# RELAY SELECTION FOR MULTIPLE ACCESS RELAY CHANNEL WITH DECODE-FORWARD AND ANALOG NETWORK CODING

Ahmed Hassan M. Hassan, Bin Dai, Benxiong Huang, Edriss Eisa and Muhammad Azhar

Department of Electronics and Information Engineering
Huazhong University of Science and Technology
Wuhan430074, P.R. China
{engahmed5577, nease.dai, bonzoga20, azharhust}@gmail.com and bxhuang@hust.edu.cn

***ABSTRACT***

*This paper presents a relay selection for decode-and-forward based on network coding (DF-NC) and analog-NC protocols in general scheme of cellular network system. In the propose scheme the two source node simultaneously transmit their own information to all the relays as well as the destination node, and then, a single relay i.e. best with a minimum symbol error rate (SER) will be selected to forward the new version of the received signal. Simulation results show that, the DF-NC scheme with considerable performance has exactness over analog-NC scheme. To improve the system performance, optimal power allocation between the two sources and the best relay is determined based on the asymptotic SER. By increasing the number of relays node, the optimum power allocation achieve better performance than asymptotic SER.*

## 1. INTRODUCTION

Cooperative communications is a specific area of wireless communication that has been extensively explored within the last decade. The concept of cooperative communications (relay network (RN)) builds upon a network architecture where nodes assist each other to realize spatial advantages of diversity. However, ever since the RN emerges, it has led to introduce new challenges i.e., low reliability, limited throughput, coverage area, service's quality and high data rate. Meanwhile, the network coding (NC) approach was established to be an efficient technique to address same challenges. Motivated by these problems this study deals with presents the analytical solutions that address these issues.

The two main familiar strategies in a wireless RN are decode-and-forward (DF) and amplify and forward (AF) [1], [2], which both schemes can be readily applied in MARC system. The benefits of DF-NC protocol were an increase in the throughput and a decrease the relay energy than the traditional scheme i.e. takes four-time slots. In contrary, the transmission in MARC with the analog-NC takes place in two-time slots. First, the two source nodes broadcast at the same time their information to one or multiple relays. The relays receive a superimposed signal, then amplify the received signal by the best relay, and forward it to destination node. Analog-NC has proven particularly useful in wireless networks due to the wireless channels' implementation of NC [3].

Recently, it has been shown that the performance of wireless RNs can be further improved by selecting an optimal relay node for transmission. In [2] and [7] the analysis of the performance for the multiple accesses relays channel (MARC) deal with single relay and not conforms to the demand for wireless networks. In this study, we consider MARC with *N*-relay to improve the spectral efficiency while one relay has higher symbol error rate (SER). We obtained the close form expression for SER and outage probability using moment generating function (MGF), probability distribution function (PDF) and cumulative distribution function (CDF) of end-to-end signal-to-noise (SNR) for i.i.d channels. The results show that the DF-NC achieved better performance than analog-NC.

Power allocation has intensively been presented in three nodes network [4]. The power allocation developed for the three nodes network cannot be able to apply directly in MARC system. This is due to different works and complexity in MARC system. In practical application of wireless RN such as a cellular system, the optimal power allocation plays a vital role for several uses. To minimize the SER of this system, the optimization of the power allocation has been investigated in this article. Hence, it is tremendously useful to study this issue, minimizing the SER for MARC. The results provide that the optimum power allocation attain better performance by increasing the number of relay.

The rest of this article is organized as follow: Section-2 introduces the system model as well as the strategies of relay selection for MARC in Section-3. The performance analysis provides in Section-4. The outage probability and power allocation are presented in Section-5. The simulation results are provided in Section-6. Finally, conclusions are drawn in Section-7.

# 2. SYSTEM MODEL

We consider a general model of MARC with analog-NC and DF-NC which consists of source nodes, denoted by $S_1$ and $S_2$, *N*-relay nodes, denoted by $R_1 \ldots R_N$ and one destination known as *D*. In the optimal relay selection scheme, as shown by the solid line in Figure 1, each message transmits from the two source nodes takes place in two modes. In the first mode, both source nodes send the information to all relays simultaneously. In the second mode, best relay node is selected to forward the received signals to destination node. For the sake of simplicity, we assume that the source and the relay nodes have all the link information. A drive an approximate formula for SER to pick the suitable relay according to the maximum SNR is presented.

In each source node the information bits $b=[b_1,b_2,...,b_k], b\in\{0,1\}$ are encoded to code bits $C=[c_1,c_2,...,c_N]$, mapped into $X=[x_1,x_2,...,x_N]$, $X\in\{\pm 1\}$ and transmitted to the relay. In this article, we restrict to BPSK for the simple presentation. The channel coefficients between sources to relays, relays to destination and sources to destination denoted as a complex value coefficients $|h_{z_1}|, |h_{z_2}|$ and $|h_{z_3}|$ where $z_1=(S_{i=1,2}, R_{j=1,...,N})$, $z_2=(R_{j=1,...,N}, D)$ and $z_3=(S_i, D)$. We assume that the channel coefficients are available at destination node. In addition, the additive white Gaussian noise on all links are $n_x, n_y$ *and* $n_z$ which are model it as normal distribution $N(0,\sigma_{z_1}^2)$, $N(0,\sigma_{z_2}^2)$ and $N(0,\sigma_{z_3}^2)$ respectively.

The received signals at relay node and destination in the first time of transmission written as:

$$y_{R_j} = \sqrt{P_{S_i}} h_{z_1} x_{S_i} + \sqrt{P_{S_2}} h_{z_2} x_{S_2} + n_x \qquad (1)$$

$$y_D^1 = \underbrace{\sqrt{P_{S_i}} h_{z_1} x_{S_i}}_{y_{S_1}^D} + \underbrace{\sqrt{P_{S_2}} h_{z_2} x_{S_2}}_{y_{S_2}^D} + n_D^1 \quad (2)$$

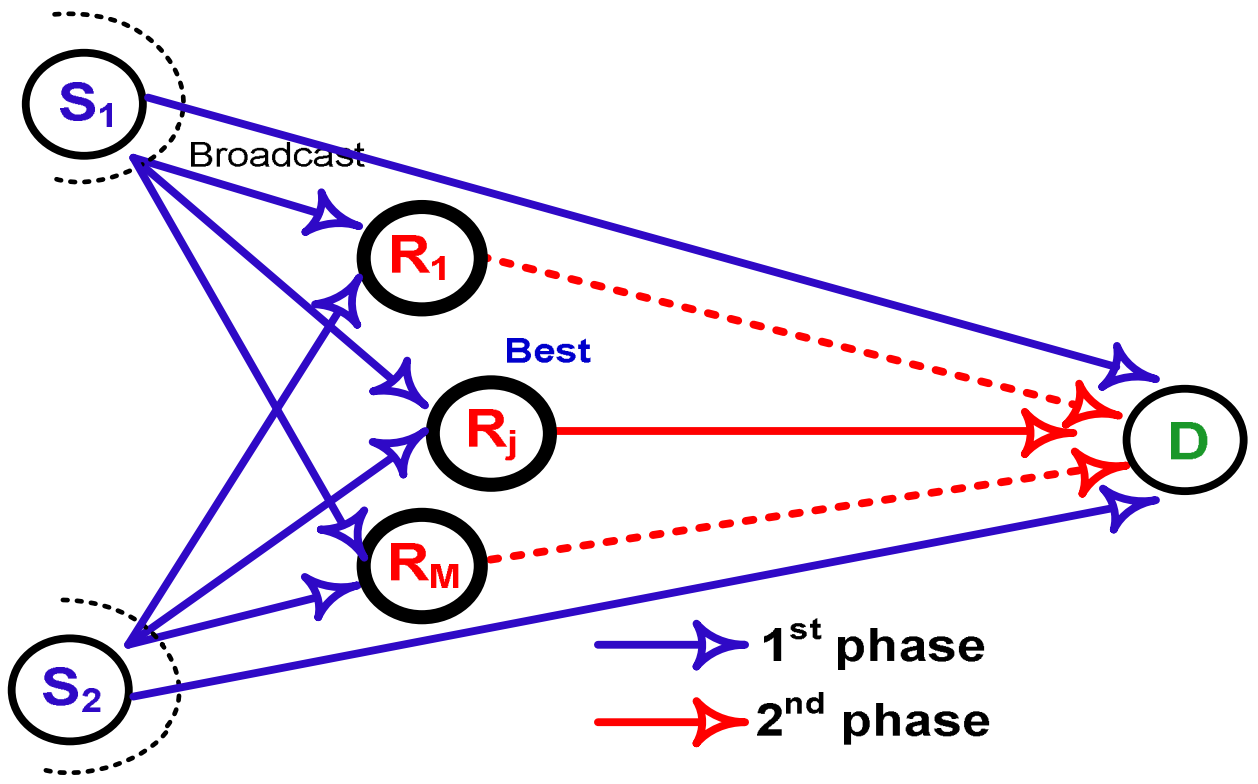

Figure 1 Block diagram of the MARC system model

The power constraint $\left(P_{S_i}\right)$ at the source nodes is $1/N\left(E[\| X_{S_i} \|^2]\right) \le P_{S_i}$ .

The output of the relay is modeled as:

$$y_D^2 = \sqrt{P_{R_j}} h_{z_2} f\left(y_{R_j}^{best}\right) + n_D^2 \quad (3)$$

where $f(.)$ denotes the process applied in the relay node. The relay amplifies the received signal and broadcasts during the second-time slot. The relay will normalize the received signal by a factor of $\beta = \sqrt{E\left[| y_R |^2\right]}$, so that the average energy is unity.

As such, this paper will focus on wireless multicast network ingrained with a two source, relays as well as destination in an effort to enhance a successful applied maximum ratio combing and receipt of information through the transmitters.

## 3. RELAY SELECTION FOR MARC

In the second-mode of transmission, just single relay is selected out of *N*-relays to forward the received signals. This is depending on the SNR of the channel link between the sources and relays node, $R_j^{best}$ may be a more suitable relay will prefer to complete the transmission. Just a single source node (i.e., $S_1$ *or* $S_2$) will determine the optimal relay according to a certain criterion and broadcast the index of it to all relays. Then, only the selected one, which is known by both source nodes, is active in the second mode of transmission, and the other keep idle. The mathematical steps of the relay selection present in the following corollary:

**Corollary:** In order to find the best relay, we can put the summarization in form as follow:

$$CDF \xrightarrow{\frac{d(F(\gamma))}{dx}} PDF \xrightarrow[0]{\int^{\infty} e^{-s\gamma} f(\gamma) d\gamma} MGF$$

$$MGF \xrightarrow{L^{-1}(MGF)} PDF \xrightarrow[0]{\int^{\infty} f(\gamma) d\gamma} CDF$$

where $F(\gamma)$ denotes the CDF, $f(\gamma)$ is PDF, and $L^{-1}(.)$ is the inverse Laplace transform.

From the *N*-relays, the one best relay is chosen as the relays which have the maximum SNR. Therefore,

$$R_j^{best} = \min_j \left\{ \overline{SER}_{S_1,j}(\gamma_{S_1,j}) + \overline{SER}_{S_2,j}(\gamma_{S_2,j}) \right\} \tag{4}$$

where $\overline{SER}_{S_i,j}$ denotes the SER at $S_1$ from the j[th] relay, $i=1,2$ and $j=1,\ldots,N$. So, the instantaneous SNR for the best relay selection is given as:

$$\gamma_{s_i,j} = \frac{|Signal|^2}{\mathrm{var}(Noise)} \tag{5}$$

# 4. Performance Analysis

We present in this section the selection methods according to two schemes (analog-NC and DF-NC) of the system model. Let us first calculate the PDF of SNR in both schemes. Due to symmetric between $\gamma_{S_1,j}$ and $\gamma_{S_2,j}$, lead to the equality of $f_{1,j}(.)$ and $F_{1,j}(.)$, denoted by PDF and CDF respectively. From equation (5), we use the $\gamma_{S_1,j}$ for derivation of both schemes, which can be written as which can be written as in [2] and [8] for analog-NC and DF schemes respectively:

$$\gamma_{1,R_j}^{ANC} = \frac{\Gamma_S |h_{z_1}|^2 \Gamma_{R_j} |h_{z_3}|^2}{|h_{z_1}|^2 \Gamma_S + |h_{z_3}|^2 \Gamma_{R_j} + 1} \tag{6}$$

$$\gamma_{1,R_j}^{DF} = |h_{z_1}|^2 \Gamma_{R_j} \tag{7}$$

where $\gamma_{1,R_j}^{ANC}$ has an exponential random variable with parameter $\eta_{1,R_j}^{ANC} = N_0 / P_{S_i}\sigma_{z_3}^2$ and $\gamma_{R_j}$ is harmonic mean of two exponential random variables which can be approximated as an exponential random variable with rate $\eta_{1,R_j}^{ANC} = 1/\Gamma_S\sigma_{z_1}^2 + 1/\Gamma_{R_j}\sigma_{z_2}^2$ at high SNR, the $\gamma_{1,R_j}^{DF}$ of the source relays links given as $\gamma_{1,R_j}^{DF} = |h_{z_3}|^2 P_{R_j}/N_0$ where $\gamma_D$ is an exponential random variable with rate $\eta_{1,R_j}^{DF} = 1/\Gamma_{R_j}\sigma_{z_3}^2$ , $\Gamma_S = P_S/N_0(1+\kappa)$ and $\Gamma_{R_j} = P_R/N_0$ . For simplicity we can assume that $P_S = \kappa P_{R_j}, \kappa > 0$.

Define $\gamma_j^{\min} \triangleq \min\{\gamma_{S_1,j}, \gamma_{S_2,j}\}$. Let $f_{\gamma_j^{\min}}(.)$ and $F_{\gamma_j^{\min}}(.)$ denotes the PDF and CDF respectively. Then the CDF of $\gamma_{R_j^{best}}$ can be calculated as:

$$\begin{aligned} F_{\gamma^{best}}(\gamma) &= P\left\{\max \sum_{i=1}^{N} \gamma_{R_j^{best}{}_i} \le \gamma \right\} \\ &= N\prod_{i=1}^{N} P\left\{\gamma_{R_j^{best}} \le \gamma\right\} = \left(F_{\gamma_{R_j^{best}}}(\gamma)\right)^N \end{aligned} \tag{8}$$

According to the system model we have that $\gamma_{R_j^{best}}$ is an exponential random variable with parameter $\eta_{R_j^{best}}$ . So the CDF is defined as:

$$F_{\gamma_{R_j^{best}}}(\gamma)=\left(1-e^{-\eta_{R_j^{best}}\gamma}\right) \tag{9}$$

Therefore substituting equation (9) into equation (8), the CDF of $\gamma_{R_j^{best}}$ can be written as:

$$F_{\gamma_{R_j^{best}}}(\gamma)=\left(1-e^{-\eta_{R_j^{best}}\gamma}\right)^N=\sum_{n=0}^{N}\binom{N}{n}(-1)^n e^{-n\eta_{R_j^{best}}\gamma} \tag{10}$$

where $\binom{N}{n}=N!/n!(N-n)!$ represents the binomial coefficient. From equation (10) we can obtain the PDF of $\gamma_{R_j^{best}}$ which is the derivative function of the CDF with respect to $\gamma$. So the PDF of $\gamma_{R_j^{best}}$ is given by:

$$f_{\gamma_{R_j^{best}}}(\gamma)=\sum_{n=1}^{N}n\eta_{R_j^{best}}\binom{N}{n}(-1)^{n-1}e^{-n\eta_{R_j^{best}}\gamma} \tag{11}$$

Using the definition of the MGF, we can define $\gamma_{R_j^{best}}$ given by using equation (11) as:

$$M_{\gamma_{R_j^{best}}}(s)=\int_0^\infty e^{-s\gamma}f_{\gamma_{R_j^{best}}}(\gamma)\,d\gamma \tag{12}$$

So substituting equation (11) into equation (12) and computing the integral we finally obtain the expression of the moment generating of $\gamma_{R_j^{best}}$ as follows:

$$M_{\gamma_{R_j^{best}}}(s)=\sum_{n=1}^{N}\binom{N}{n}n(-1)^{n-1}\left(\frac{\eta_{R_j^{best}}}{\left(s+\eta_{R_j^{best}}\right)}\right) \tag{13}$$

Using general equation of average SER [5], the SER from relay selection can be expressed as:

$$\overline{SER}_{R_j}=\int_0^{(M-1)\pi/M}M_{\gamma_{R_j^{best}}}(-s)\left(\frac{g}{\sin^2\theta}\right)d\theta \tag{14}$$

where g =1 for BPSK as well as $M=2$ and MGF $M_{\gamma_{R_j^{best}}}$ given by equation (13).The average SER (in case of BPSK it known as BER) for our scheme is given by averaging equation (14) and Substitute the different moment generating functions by their values, hence the expression becomes:

$$\overline{SER}_{R_j}=\sum_{n=1}^{N}\binom{N}{n}n(-1)^{n-1}\times\frac{1}{\pi}\int_0^{\pi/2}(A)(B)\,d\theta \tag{15}$$

where $A=\left(\frac{\sin^2\theta}{n\sin^2\theta+c_1}\right)$, $B=\left(\frac{\sin^2\theta}{\sin^2\theta+c_2}\right)$, $c_1=b/\eta_{R_j}$, $c_2=b/\eta_{S_i,D}$, $b=\sin^2(b/M)$ .The expression in equation (15) represents the average SER from relay selection. Using partial fraction expansions and after some manipulations it can follows that [6]:

$$\overline{SER_{R_j}} = \sum_{n=1}^{N} \binom{N}{n} (-1)^{n-1} \left( I(c_1) + I(c_2) \right) \quad (16)$$

$$where, I(c_e)\,|_{e=1,2} = \frac{1}{\pi} \int_0^{\pi/2} \frac{\sin^2\theta}{\sin^2\theta + c} d\theta = 0.5\left[1 - \sqrt{\frac{c}{1+c}}\right]$$

# 5. Outage Probability and Power Allocation

In this section, we provide how we can calculate the outage probability and power allocation to both schemes.

## 5.1 Outage probability

Outage probability defined as the probability that the output SNR falls a certain given threshold, mathematically it is:

$$P_{out} = \int_0^{\gamma_{th}} f_{\gamma_{R_j^{best}}}(\gamma) d\gamma \quad (17)$$

where $f_{\gamma_{R_j^{best}}}(\gamma)$ denotes the PDF of SNR. Therefore, using the expression in (11), the Pout written as:

$$P_{out} = \sum_{n=1}^{N} n\eta_{R_j^{best}} \binom{N}{n} (-1)^{n-1} \int_0^{\gamma} e^{-n\eta_{R_j^{best}}\gamma} d\gamma \quad (18)$$

## 5.2 Power allocation

Power allocation has been investigated to minimize the outage probability and maximize the mutual information of the analog-NC scheme [4]. In this work, we can provide here the power allocation to both sources and the SER best relay that has lower bound i.e. $0 \le sin^2\theta \le 1$ of subject to total power transmission.

$$\begin{aligned} &min \quad \left(\overline{SER}_{best}\right) \\ &subject\ to \\ &\quad \left(2P_s + P_r = P_{total}\right) \end{aligned} \quad (19)$$

The optimum power allocation is to find the source power that minimizes the lower bound of SER subject to the total power by solving the *Lagrange* equation below:

$$\Omega(P_s) = SER_{R_j} + \Delta(2Ps + P_R - P_{total}) \quad (20)$$

where $\Delta$ is positive Lagrange multiplier. By driving equation (20) with respect to $P_s$ and $P_R$ equal to zero, respectively. We can get:

$$\frac{\partial \Omega(P_s)}{\partial P_s} = \frac{\partial SER_{R_j}}{\partial P_s} + 2\Delta = 0$$
$$\frac{\partial \Omega(P_{R_j})}{\partial P_{R_j}} = \frac{\partial SER_{R_j}}{\partial P_{R_j}} + \Delta = 0 \qquad (21)$$

By integrating the $2P_s + P_r = P_{total}$ and $SER_{R_j}$ given in equations (16) and (21), we can obtain:

$$P_s = (1/(4b))(A+(B/A)+C)$$
$$P_{R_j} = 2Ps - P_{total} \qquad (22)$$

where,

$$A = (-135P_{total}b - 9P_{total}{}^2b^2 + 91P_{total}{}^3b^3 - 27 + 12P_{total}$$
$$sqrt(-486P_{total}b - 756P_{total}{}^2b^2 - 81 - 402P_{total}{}^3b^3 - 51P_{total}b^4)b)^{(1/3)}$$

$B = 30P_{total}b + 25P_{total}^2b^2 + 9$ and $C = (-3+7Pb)/b$

which indicate the power at the relay should be equal to the total power at the two sources to indemnity for the energy used generates the new version of combined information in one time slot, irrespective of number of relays.

## 6. Simulation Results

In this section, we provide the simulation results using MATLAB tools for the MARC model. Due to symmetrical sources, they should have the same SER; therefore, it would be satisfy to examine one source. The simulation results are provided for a BPSK modulation over the Rayleigh fading channels. We present the simulation results of the closed form of approximate SER to pick a best relay among *nR*-relays node of Analog-NC and DF-NC strategies.

Figure 2 show the average of SER of the system model with BPSK and 8PSK for *N*-relays. We can observe that in both cases the average SER inversely proportional to number of relay (i.e. $\overline{SER}$ decrease when the number of relay increase). The compression between analog-NC and DF scheme are available in Figure 3. It easy to observe that the DF protocol is outperforms the Analog-NC. The outage probability versus SNR results of the system model presented in Figure 4. We can observe from it that the number of relay nodes effect the outage probability (i.e. decrease).

In Figure 5, we examine the performance of SER of analog-NC scheme with reference value of $P_s$ and $P_r$ given by equation (22). The asymptotic results are provide for comparison with equal power $P_s = P_{Rj} = P_{total}$. We can easily observe that, the SER with optimum power allocation achieved better performance than an asymptotic scheme even the number of relays node are increased.

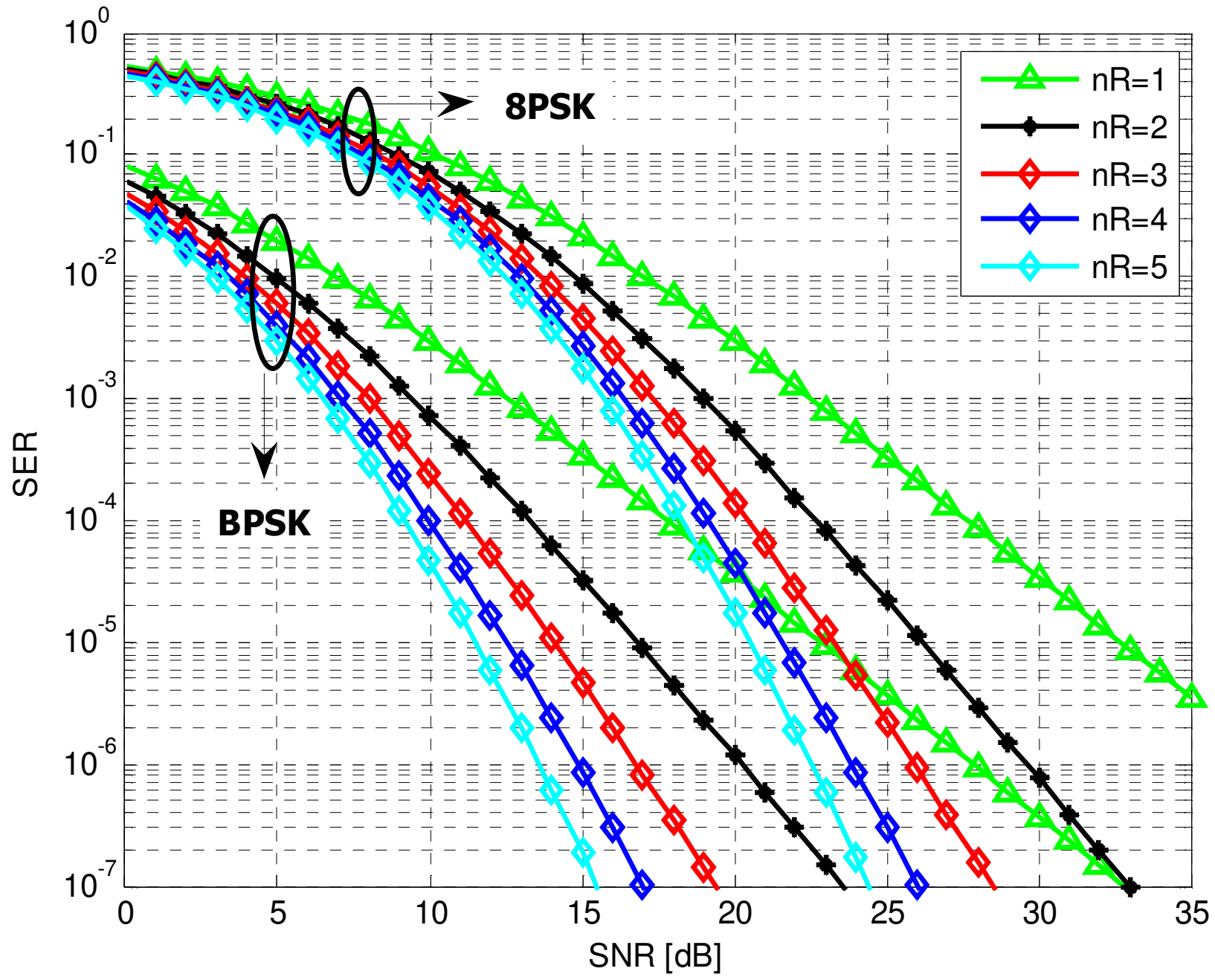

Figure 2 SER versus SNR for BPSK and 8-PSK in case of analog-NC, where nR = 1, 2, 3, 4, 5.

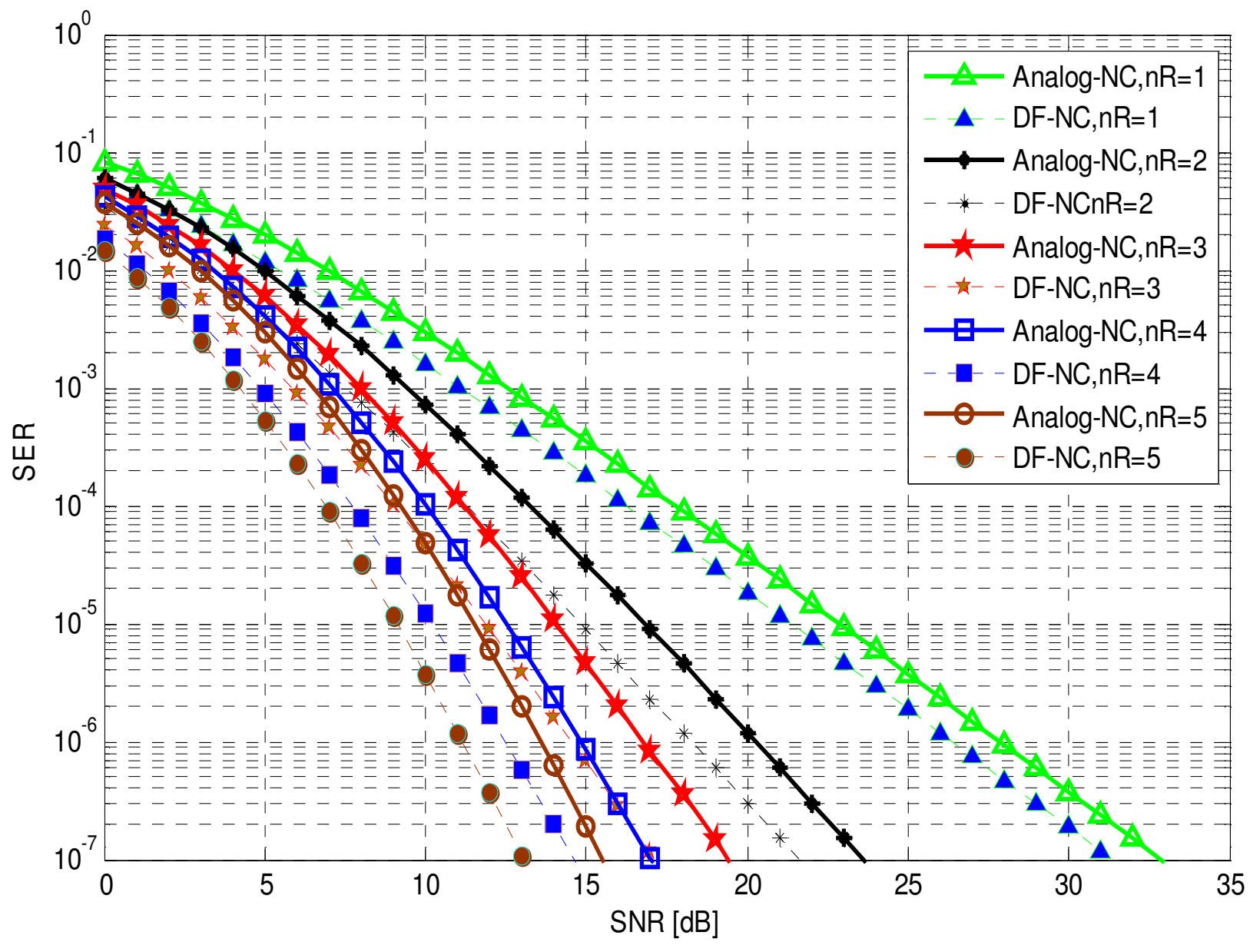

Figure 3 Comparison of SER versus SNR for BPSK in case of Analog-NC and DF protocol, where nR = 1, 2, 5, 10

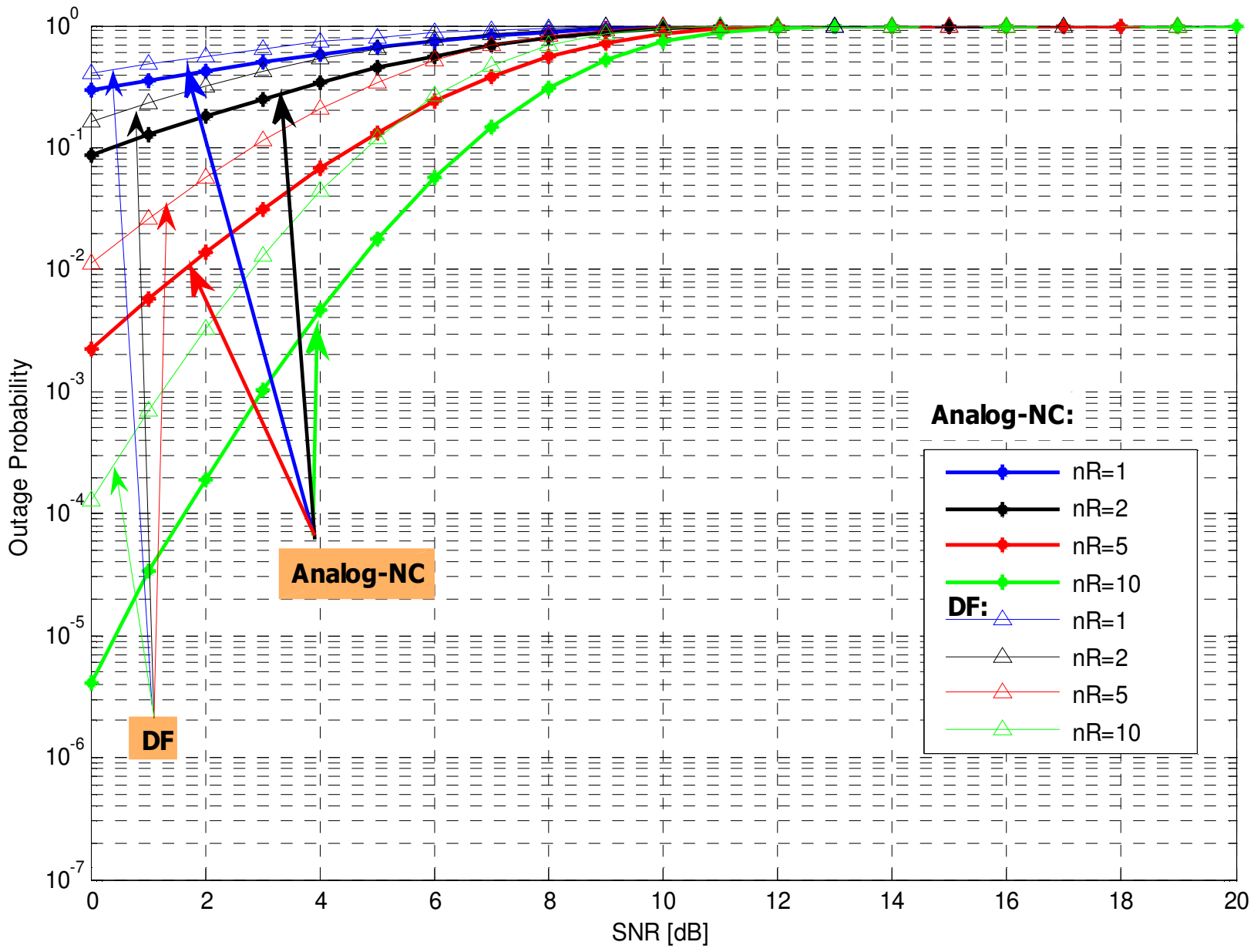

Figure 4 Comparison of Outage Probability versus SNR [dB] in case of Analog-NC and DF protocol, where nR = 1, 2, 5, 10

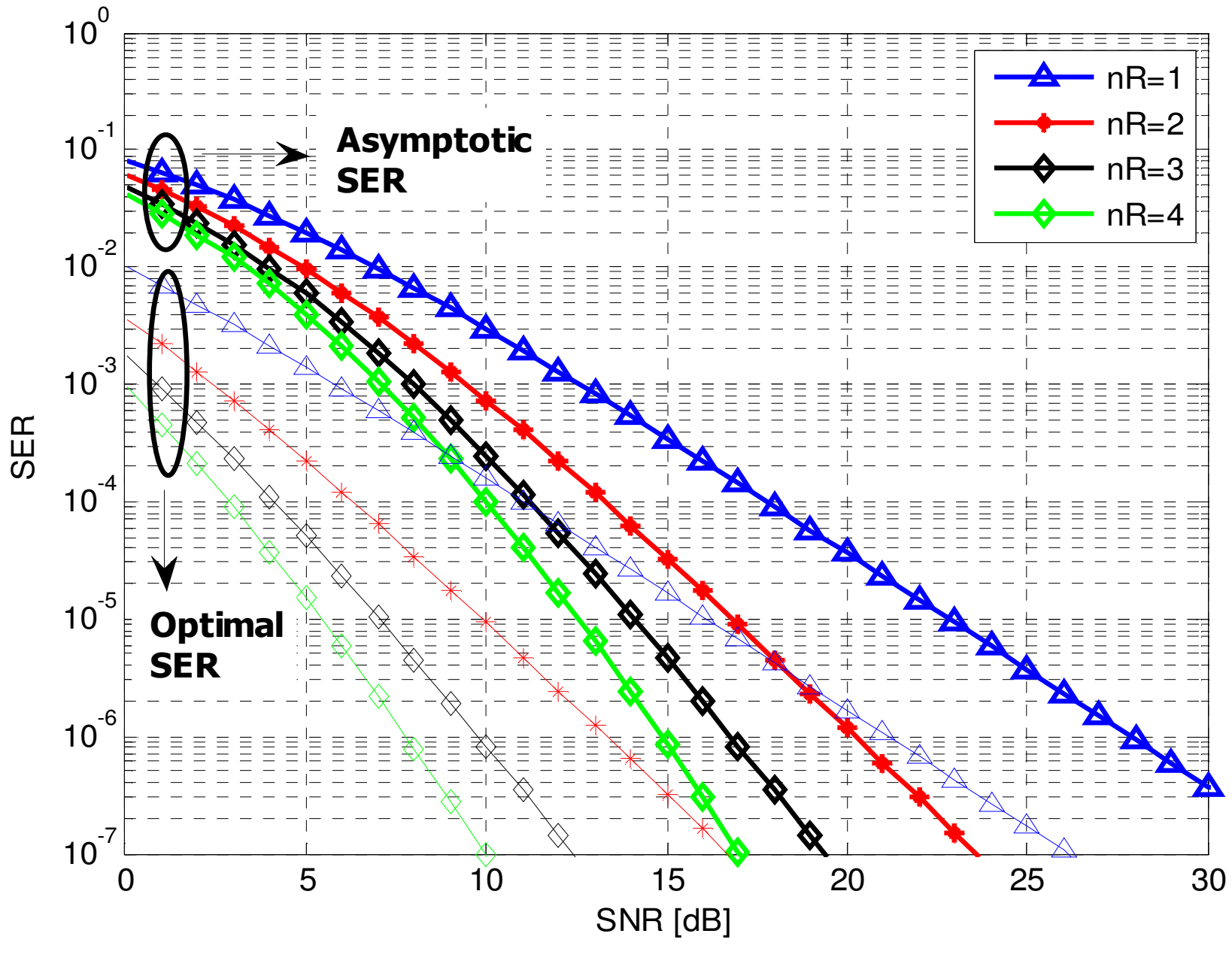

Figure 5 Simulated the SER by analog-NC scheme using transmit power allocation, where nR =1, 2, 3, 4.

# 7. Conclusion

In this wok, we have proposed a relay selection with DF-NC and analog-NC over MARC system. We derived the asymptotic close forms of SER for both schemes, which are verified through simulation. We demonstrated by analysis, the DF-NC scheme has good performance over a wide range of SNR than the analog-NC. To improve the system performance, the optimum power allocation between the two sources and the best relay is provided based on the asymptotic SER. The analytical results proved that the optimum power allocation has considerable performance over asymptotic SER, mainly using a large number of relays node.

## ACKNOWLEDGMENT

This work was supported by the National Natural Science Foundation of China under Grant No.60803005 and the Important National Science and Technology Specific Projects 2009ZX03004-004 and 2010ZX03003-003.Bin Dai is the corresponding author.

**Authors:**

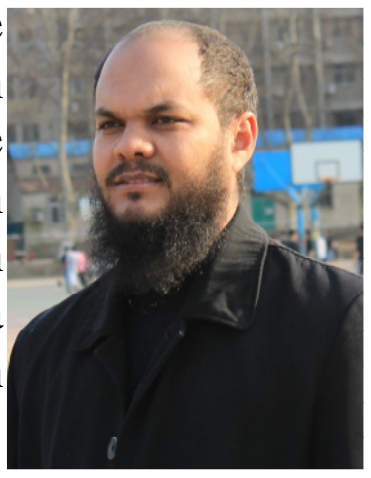

**Ahmed Hassan** received his B.Eng. degree in Electrical Engineering from the Blue Nile University, Sudan, in 2003 and his Master degree in Electronics and Information Engineering from the Sudan University of Science and Technology, Sudan, in 2006. He is currently a Ph.D. candidate at the Department of Electronics and Information Engineering in Huazhong University of Science and Technology, China. His research interests include wireless Network Coding and cooperative networks. He was a recipient of awards from China's scholarship Council (CSC) at 2008 in cooperation with Sudan's government, and he is an IEEE member.
E-mail address: engahmed5577@gmail.com

**Bin Dai** received the B. Eng, the M. Eng degrees and the PhD degree from Huazhong University of Science and Technology of China, P. R. China in 2000, 2002 and 2006, respectively. From 2007 to 2008, he was a Research Fellow at the City University of Hong Kong. He is currently an associate professor at Department of Electronic and Information Engineering, Huazhong University of Science and Technology, P. R. China. His research interests include p2p network, wireless network, network coding, and multicast routing.
Email address: nease.dai@gmail.com

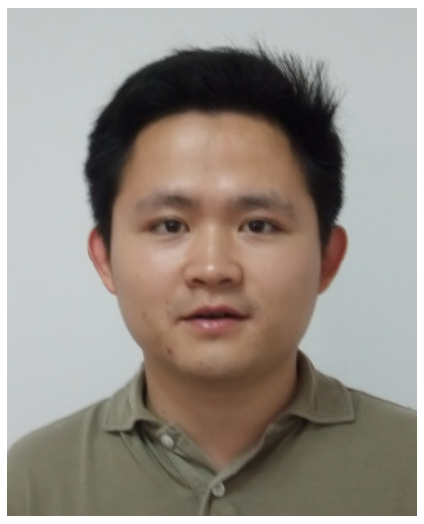

**Benxiong Huang** received the BS.C in 1987 and PhD in 2003 from Huazhong University of Science and Technology (HUST), P.R China. He is currently professor in the Department of Electronic and Information Engineering at HUST. His research interests include next generation communication system and communication signal processing.
Email address: bxhuang@hust.edu.cn

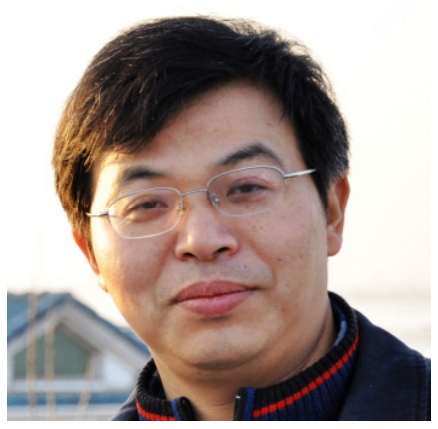

**Edriss Eisa** received his B.S. degree in electrical engineering from Blue Nile University, Damazein, Sudan, in 2004 and his Msc. degree in Electronics and Information Engineering from Sudan University of Science and Technology, Sudan, in 2008. He is currently working towards the Ph.D. degree at the Department of Information Engineering in Huazhong University of Science and Technology, China. His current research interests are in the areas of cooperative wireless communications, and sensor MIMO systems. Mr. Adam received CHINASCHOLARSHIP COUNCIL CSC 2010 cooperation with Sudan government.
E-mail address: bonzoga20@gmail.com

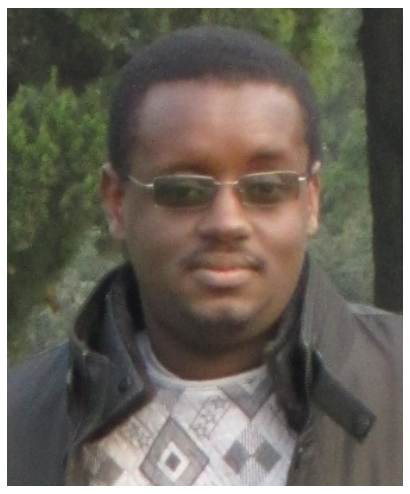

**Muhammad Azhar** is Ph.D. student at the Department of Information Engineering in Huazhong University of Science and Technology, China. His current research interests in the areas of vehicular ad hoc network and network coding.
E-mail address: azharhust@gmail.com

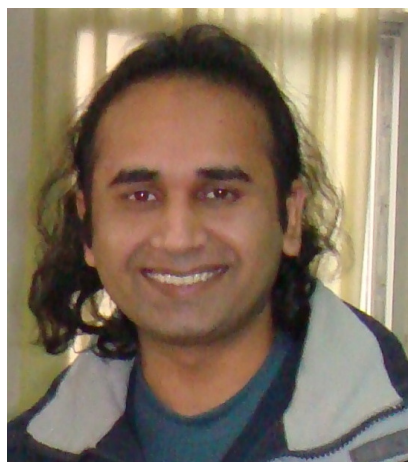